\documentclass[aps,prl,reprint,modulo,showpacs,groupedaddress,amssymb]{revtex4-1}
\pdfoutput=1 
\usepackage{atlasphysics}
\usepackage[hang,center]{subfigure}
\usepackage{dcolumn}
\usepackage{xspace}
\usepackage[pdftex]{graphicx}
\usepackage{color}
\usepackage{preprintcover}
\PassOptionsToPackage{hyphens}{url}
\usepackage{hyperref}
\usepackage{amsmath}

\newcommand{\pythia}{{PYTHIA}}
\newcommand{\herwig}{{HERWIG}}
\newcommand{\jimmy}{{JIMMY}}

\newcommand{\sherpa}{{SHERPA}}

\newcommand{\acermc}{{AcerMC}}
\newcommand{\alpgen}{{ALPGEN}}
\newcommand{\qbh}{{QBH}}
\newcommand{\mcnlo}{{MC@NLO}}
\newcommand{\mcnlonospace}{{MC@NLO}}
\def\BFCS{\ensuremath{\Sigma \sigma_{\mathrm{qq}} \times \mathrm{BF_{qq}}}}

\PreprintCoverPaperTitle{Search for Quantum Black Hole Production in High-Invariant-Mass Lepton+Jet Final States Using $pp$ Collisions at $\sqrt{s} = 8$\,\tev\ and the ATLAS Detector} 

\PreprintIdNumber{CERN-PH-EP-2013-193}  

\PreprintCoverAbstract{
This Letter presents a search for  quantum black-hole production using $20.3$\,fb$^{-1}$ of data collected  with the ATLAS detector in $pp$ collisions at the LHC at $\sqrt{s} = 8$\,\tev.  The quantum black holes are assumed to decay into a final state characterized by a lepton (electron or muon) and a jet.  In either channel, no event with a lepton--jet invariant mass of $3.5$\,\tev\  
or more is observed, consistent with the expected  background. Limits are set on the product of cross 
sections and branching fractions for the lepton+jet final states of quantum black holes produced in a search region  for invariant masses above $1$\,\tev.  
The combined $95$\% confidence level upper limit on this product for quantum black holes with threshold mass above $3.5$\,\tev\ 
is $0.18$\,fb.  This limit constrains the threshold quantum black-hole mass to be above $5.3$\,\tev\ in the model considered.
}

\PreprintJournalName{Physical Review Letters}  

\begin{document}
\title{ Search for Quantum Black Hole Production in High-Invariant-Mass Lepton+Jet Final States Using $pp$ Collisions at $\sqrt{s} = 8$\,\tev\ and the ATLAS Detector}
\author{The ATLAS Collaboration}
\begin{abstract}\noindent
This Letter presents a search for  quantum black-hole production using $20.3$\,fb$^{-1}$ of data collected  with the ATLAS detector in $pp$ collisions at the LHC at $\sqrt{s} = 8$\,\tev.  The quantum black holes are assumed to decay into a final state characterized by a lepton (electron or muon) and a jet.  In either channel, no event with a lepton--jet invariant mass of $3.5$\,\tev\  
or more is observed, consistent with the expected  background. Limits are set on the product of cross 
sections and branching fractions for the lepton+jet final states of quantum black holes produced in a search region  for invariant masses above $1$\,\tev.  
The combined $95$\% confidence level upper limit on this product for quantum black holes with threshold mass above $3.5$\,\tev\ 
is $0.18$\,fb.  This limit constrains the threshold quantum black-hole mass to be above $5.3$\,\tev\ in the model considered.
\end{abstract}

\maketitle

Quantum black holes (QBHs)~\cite{Meade:2008,Calmet:2008} are predicted in  low-scale quantum gravity theories that offer 
solutions to the mass hierarchy problem of the Standard Model (SM) by lowering the scale of quantum gravity (\md) from 
the Planck scale ($\sim 10^{16}$\,\tev) to a value of about $1$\,\tev.  In models with large extra 
dimensions such as the Arkani-Hamed--Dimopoulous--Dvali 
(ADD) model~\cite{ArkaniHamed:1998rs, Antoniadis:1998ig, ArkaniHamed:1998nn}, only the gravitational field is allowed to penetrate 
the $n$ extra dimensions, while all SM fields are localized in the usual four-dimensional space-time.  QBHs with masses near \md, postulated to conserve total angular momentum,  color and electric charge, may decay to two particles~\cite{gingrich:BHpaper,Calmet:2008}.  The behavior of QBHs is distinct from semi-classical black holes that decay via Hawking radiation to a large number of objects~\cite{Anchordoqui:2001cg}.  Searches for semi-classical black holes typically require three or more objects~\cite{Aad:2012ic,Chatrchyan:2013taa}.  

The quantum approximations used in the modeling of black hole production are valid when black-hole masses are above a minimal threshold mass, \mth, which is taken to be equivalent to the QBH inverse gravitational radius.  If the QBHs investigated in this Letter are accessible at the Large Hadron Collider (LHC)~\cite{LHCref}, they can produce lepton+jet final states~\cite{,Calmet:2008,gingrich:BHpaper} motivating this first dedicated search for high-invariant-mass final states with a single electron ($e$) or a single muon ($\mu$), and at least one jet.  Two-particle QBH decays to a final state consisting of a lepton and a quark-jet violate lepton and baryon number conservation, producing a distinctive signal for physics beyond the SM.  Previous searches for QBHs relied on 
signatures such as dijet mass distributions~\cite{ATLAS:2012pu,CMS:2012yf}, generic multiobject configurations~\cite{Chatrchyan:2013taa} and photon+jet final states~\cite{ATLAS:gammajetpaper}.

The largest QBH cross section for the final states considered is predicted for the collision of two $u$-quarks ($\sigma_{{uu}}$),  which produces charge $+4/3$ objects with equal branching fractions 
(BFs) of $\mathrm{BF}_{uu} = 11$\% to each lepton+jet final state.  The next largest cross sections are for charge $+1/3$  ($ud$) and $-2/3$  
($dd$) QBHs with lepton+jet BFs of $\mathrm{BF}_{ud} = 5.7$\% and $\mathrm{BF}_{dd} = 6.7$\%~\cite{gingrich:BHpaper}. Processes with initial states having antiquarks and heavier
 sea-quarks are suppressed by at least a factor of $100$ and can be neglected.   The  QBH cross section is a steeply declining function of \mth, and has $\BFCS \approx 8.6\times10^5$\,fb, $8.9\times10^2$\,fb and  $0.75$\,fb for $\mth$ of $1$\,\tev, $3$\,\tev\ and $5$\,\tev, respectively~\cite{Gingrich:2009da}.

The ATLAS detector~\cite{Aad:2008zzm} includes an inner tracker, covering a pseudorapidity~\footnote{ATLAS uses a right-handed coordinate system with its origin 
at the nominal interaction point (IP) in the center of the detector and the $z$-axis along the beam pipe.   
The $x$-axis points from the IP to the center of the LHC ring, and the $y$-axis points upward.  
The pseudorapidity is defined in terms of the polar angle $\theta$ as $\eta=-\ln\tan(\theta/2)$.  Cylindrical coordinates $(r,\phi)$ are used in the transverse plane, $\phi$ being the azimuthal angle around the beam pipe, referred to the $x$-axis. 
} range $|\eta| < 2.5$, surrounded 
by a superconducting solenoid providing a $2$\,T central field.  A liquid-argon (LAr) electromagnetic (EM) sampling calorimeter ($|\eta| < 3.2$), a scintillator-tile hadronic calorimeter ($|\eta| < 1.7$), a LAr hadronic calorimeter ($1.4 < |\eta| < 3.2$), and a LAr forward 
calorimeter ($3.1 < |\eta| < 4.9)$ provide the energy measurements. The muon spectrometer consists of tracking chambers covering 
$|\eta| < 2.7$, and trigger chambers covering $|\eta| < 2.4$, in a magnetic field produced by a system of air-core toroids.  Events considered in this analysis 
are required to have one high-transverse-momentum (high-\pt) lepton ($e/\mu$) that passes requirements of the three-level trigger 
system~\cite{Aad:2012xs}.  The thresholds applied at the third trigger level are $60$\,\gev\ and $36$\,\gev\ for electrons and muons, respectively.  The analysis is based on the complete $2012$ data set of $pp$ collisions taken at a center-of-mass energy of $\sqrt{s}=8$\,\tev\ by the ATLAS detector at the LHC, corresponding to an integrated luminosity of $20.3\pm0.6$\,\ifb~\cite{Aad:2013lumi} after data-quality requirements.

The event selection is designed to be efficient for generic lepton+jet final states and is based on leading-order 
simulated-signal QBH events obtained from the \qbh1.04 generator~\cite{Gingrich:2009da}, followed by parton 
showering and hadronization using \pythia8.165~\cite{Pythia:2007}. The signal generator uses the MSTW2008LO~\cite{Martin:2009iq} set of leading-order 
parton distribution functions (PDFs) with the AU2 underlying-event tune~\cite{atlas:pythia8tunes}.  
This Letter assumes the ADD model with $\mth=\md, n=6$, and the QCD factorization scale for the PDFs set to the inverse gravitational radius~\cite{Gingrich:2009da}.  
Samples with  $\mth$ from $1$\,\tev\ to $6$\,\tev, in steps of $0.5$\,\tev, are generated for both channels. 
 
Events with a high-\pt\ lepton and one or more jets can also arise from electroweak (EW) processes including vector-boson production
with additional jets; diboson ($WW,WZ,ZZ$), top-quark pair ($t\bar{t}$) and single top-quark ($t$ or $\bar{t}$)  production; and multijet processes including 
non-prompt leptons from semileptonic hadron decays and  jets misidentified as leptons.

The EW background in the signal region (SR) is estimated using Monte Carlo (MC) samples normalized to data in control regions.
All MC simulated samples are produced using the ATLAS detector simulation~\cite{ATLASSimInfra} based on GEANT4~\cite{Agostinelli2003250}.  The simulated events are reconstructed in the same manner as the data. The \ttbar\ and single-top-quark events are simulated with \mcnlonospace4.06~\cite{Frixione:2008yi} and \acermc3.8~\cite{Kersevan:2004yg}, respectively; the production of $W$+jets  and $Z$+jets is simulated using \alpgen2.14~\cite{Mangano:2002ea}; and 
diboson production is simulated with \sherpa1.4.1~\cite{Gleisberg:2008ta}.  The leading-order CTEQ6L1 
PDFs~\cite{Pumplin:2002vw} are used for \alpgen\ and \acermc\ samples while the next-to-leading-order CT10 PDFs~\cite{Lai:2010vv} are 
used for the \sherpa\ and \mcnlo\ samples.  The generators for all samples except dibosons are interfaced to 
\herwig6.520~\cite{Corcella:2000bw,ATLAS-PHYS-PUB-2011-008} for parton showering and hadronization and to \jimmy4.31~\cite{Butterworth:1996zw} for the underlying-event 
model.  The results of higher-order calculations are used to adjust the relative fractions of the simulated events as 
in Refs.~\cite{KfactorNNLO,KfactorNNLOTwo}.  Additional inelastic $pp$ interactions, termed pileup, are included in the event simulation so
 as to match the distribution in the data (on average $21$ interactions per bunch crossing).  

Electron candidates are identified as localized depositions of energy in the EM calorimeter with ${\pt}_e>130$\,\gev\ and $|\eta| < 2.47$, 
excluding the barrel--endcap transition region, $1.37  < |\eta | < 1.52$,  and matched to a track reconstructed in the tracking 
detectors.  Background from jets is reduced by requiring that the shower profiles are consistent with those of 
electrons~\cite{Aad:2011mk}.  Isolated electrons are selected by requiring the transverse energy deposited in a cone of 
radius $\Delta R = \sqrt{ (\Delta \eta)^2 +   (\Delta \phi)^2 }  = 0.3$ centered on the electron cluster, excluding the energy of the 
electron cluster itself, to be less than $0.0055\times{\pt}_e + 3.5$\,\gev\ after corrections for energy due to pileup and energy leakage from the electron cluster into the cone.   This criterion provides nearly constant selection efficiency for signal events over the entire ${\pt}_e$ range explored and for all the pileup conditions.

Muon candidates are required to be detected in at least three layers of the muon spectrometer and to have ${\pt}_\mu>130$\,\gev\ and $|\eta| < 2.4$.  
Possible background from cosmic rays is reduced by requiring the transverse and longitudinal distances of closest 
approach to the interaction point to be smaller than $0.2$\,mm and $1.0$\,mm, respectively.  Signal muons are required to be isolated such 
that $\sum \pt <  0.05\times{\pt}_\mu$, where $\sum \pT$ is the sum of the \pt\ of the other tracks in a cone of radius $\Delta R = 0.3$ 
around the direction of the muon.

Jets are constructed from three-dimensional noise-suppressed clusters of calorimeter cells using the anti-$k_t$ algorithm with a radius parameter 
of $0.4$~\cite{Cacciari:2008gp,Cacciari:2005hq}.  Jet energies are corrected for losses in material in front of the active calorimeter 
layers, detector inhomogeneities, the non-compensating nature of the calorimeter, and 
pileup.  Jet energies are calibrated using MC 
simulation and the combination of several in-situ techniques applied to data~\cite{2010JES, Cacciari:2008gn,Cacciari:2009dp}.  All jets are required to have ${\pt}_j > 50$\,\gev\ and 
$|\eta| < 2.5$.  In addition, the most energetic jet is required to have ${\pt}_j > 130$\,\gev.

The missing transverse momentum (with magnitude \MET), used only in the background estimation, is calculated as the negative of 
the vectorial sum of calibrated clustered energy deposits in the calorimeters, and is corrected for the momenta of any reconstructed muons~\cite{Aad:2012re}.

In the electron (muon) channel, events are required to have exactly one electron (muon).  Multijet background can be reduced, with minimal loss in signal efficiency, by requiring the average value of $\eta$ for the lepton and leading jet to satisfy $ |\langle \eta \rangle | < 1.25$ and the difference between the lepton and leading jet $\eta$ to satisfy $|\Delta \eta| < 1.5$.  The signal lepton and jet are mostly back-to-back in $\phi$ and are required to satisfy $|\Delta \phi| > \pi/2$. 

The invariant  mass (\minv) is calculated from  the lepton and highest-\pt\ jet.   
The SR is defined by a lower bound on \minv,  $\mmin$, that accounts 
for experimental resolution.  
In the electron channel $\mmin = 0.9$\mth is used.
In the muon channel, the requirement is loosened 
at high invariant mass, as muon resolution has a term quadratic in ${\pt}_\mu$, resulting in  
$\mmin=\left[0.95-0.05\mth/1\mathrm{TeV}\right]\mth$.  A low-invariant-mass control region (LIMCR) is defined with \minv\ 
between $400$\,\gev\ and $900$\,\gev, which has a negligible contamination from a potential signal ($<2\%$) for the lowest \mth\ considered. 

The acceptance of the event selection is about $65$\%, based on generator-level quantities and calculated by imposing selection criteria that apply directly to phase space (lepton/jet $\eta$, lepton/jet \pt, $ \Delta\eta$, $\Delta\phi$, $\langle \eta \rangle$, and \minv). All other selection criteria,
which in general correspond to event and object quality requirements, are used to calculate the experimental efficiency based on the events included in the acceptance.  The experimental efficiency falls from $89(59)$\% to $81(50)$\% for masses from $1$\,\tev\ to $6$\,\tev\ in the electron (muon) channel.  The experimental efficiency in the muon channel is lower than that in the electron channel because more stringent requirements are applied to ensure the best possible resolution on \minv.  The cumulative signal efficiency is the product of the acceptance and experimental efficiency.

In the electron channel, the multijet background is characterized by small values of  \MET, while EW background events can have large \MET\  due to the production of high-momentum neutrinos.  The discriminating power of \MET\ is 
used to determine the normalization of the two backgrounds to the data in the LIMCR.
The multijet template is taken from data in which electron candidates pass relaxed identification criteria but fail the normal identification selection, and  the EW template is taken from MC simulation where electron candidates pass the normal selection.  The templates are fit to the \MET\ distribution in the interval 
$[0,150]$\,\gev\ in five separate detector-motivated regions of $\eta$, to determine normalization factors for both the multijet and EW 
backgrounds.  

To extrapolate both the multijet and EW background to the SR, 
functions of the form $p_1 x^{p_2 + p_3 \ln(x) } (1 - x)^{p_4}$ (with $x = \minv/\sqrt{s}$ and fit parameters $p_1$--$p_4$)~\cite{RMHarris} 
are used and the contributions are scaled by the corresponding normalization factor derived in the LIMCR.  A simple power-law fit, 
with $p_3$ and  $p_4 $ fixed to zero, adequately describes both data and simulation.  This is used as the baseline, while $p_3$ and 
$p_4$ are allowed to vary as part of the evaluation of the systematic uncertainty.  

 In the muon channel, the multijet and EW backgrounds can be discriminated on the basis of the transverse impact parameter (\dzero) distribution of the muon since multijet background is dominated by jets containing charm and bottom hadrons decaying to muons while EW backgrounds are dominated by prompt muons.
The template for the EW background is selected using $Z$-boson decays to two muons while the template for multijet background is taken from muons that fail the isolation requirement.  Both templates are taken from data.  The templates are fit to the \dzero\ distribution in the interval $[-0.1,+0.1]$ mm to determine the normalization factors. The fraction of multijet background, $0.046 \pm 0.005$, is neglected when extrapolating the background in SR.  The procedure for extrapolating the EW background to the SR is the same as for the electron channel.
\begin{figure}[t]
  \includegraphics[width=\columnwidth,trim=4 0 4 4,clip=true]{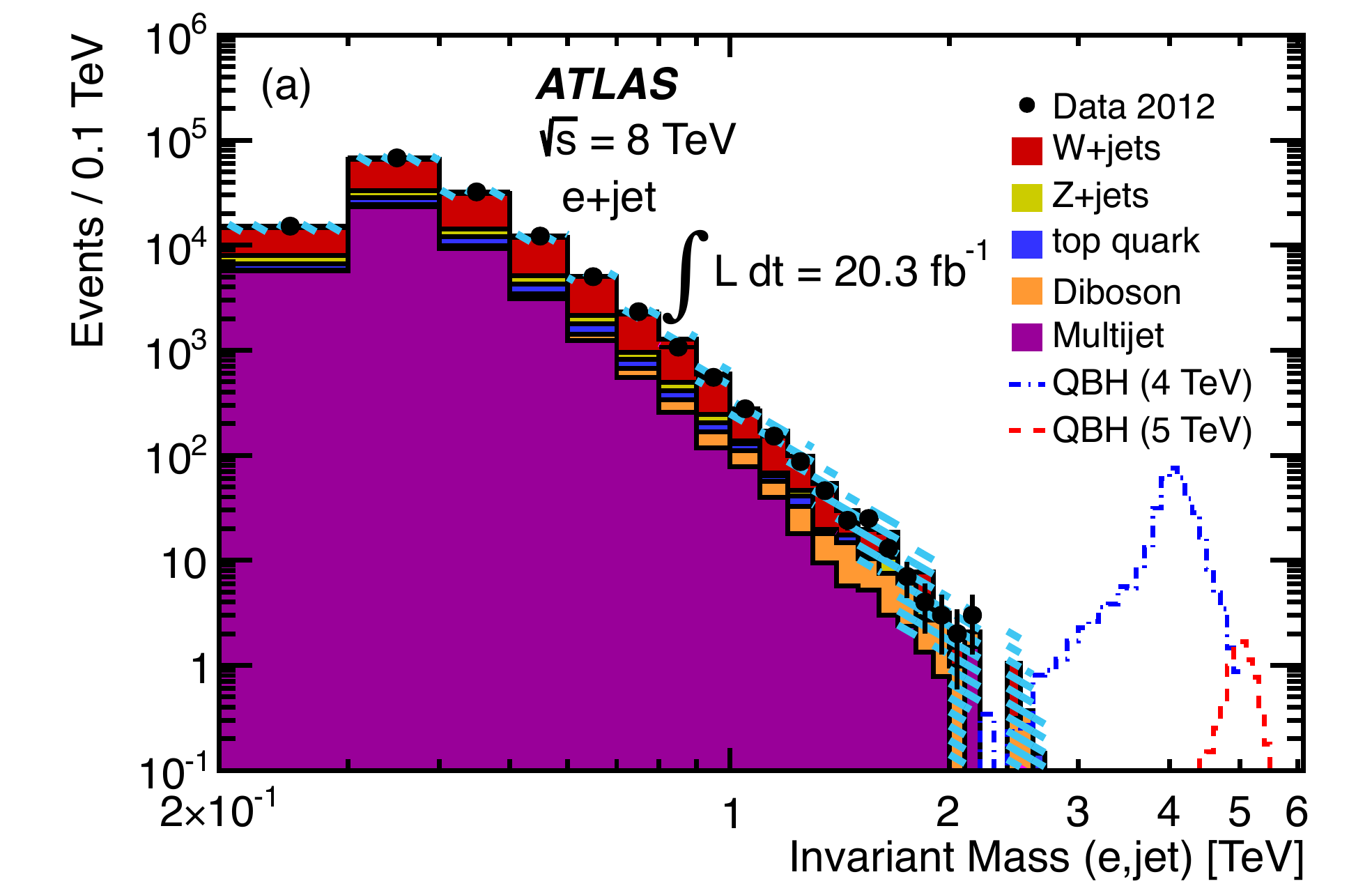}
  \includegraphics[width=\columnwidth,trim=4 0 4 4,clip=true]{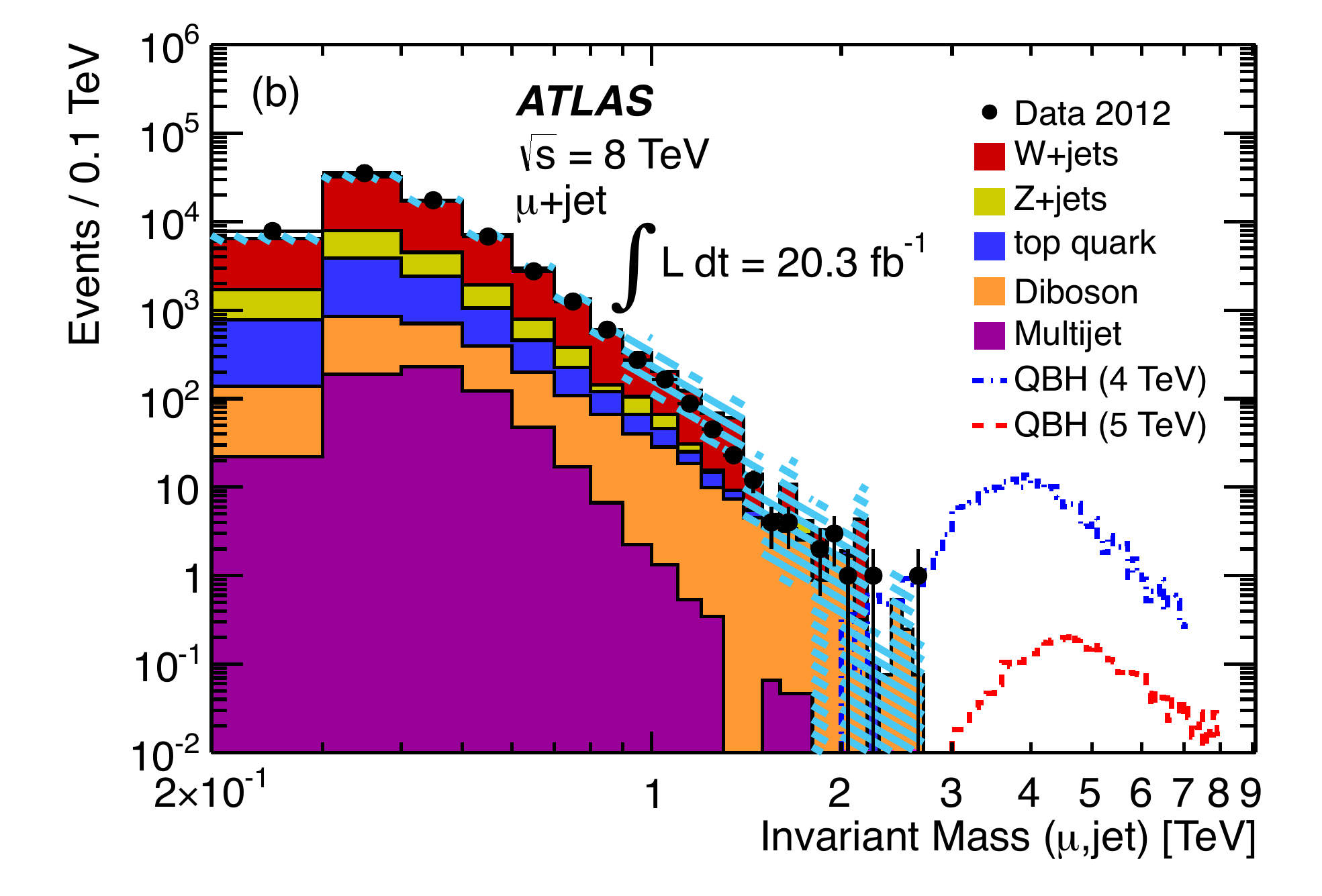}
  \caption{\label{fig:minv} Distribution of the invariant mass of the lepton and highest-\pt\ jet in (a) the electron+jet channel and (b) the muon+jet channel, for data (points with error bars) and for SM backgrounds (solid histograms).   Overlaid are two examples of QBH signals.    
    The sum of the uncertainties due to the finite MC sample size and from various sources of systematic uncertainty is shown by the hatched area.
    To extract the upper limit on the lepton+jet cross section, a fit to the invariant-mass distribution is performed, replacing the uncertainties due to MC sample size by the statistical uncertainties on the fit parameters.}
\end{figure}

The background estimate in the SR, shown in  Fig.~\ref{fig:minv}, was not compared to data until the final fit method and parameters were fixed. The hatched area in Fig.~\ref{fig:minv} shows the total uncertainty 
in the background estimate, which is dominated by the systematic uncertainties.  In extracting the limits, the fits described above
are used to extrapolate the background into the high invariant-mass region.
\begin{table}[t]
  \caption{
    Breakdown of relative systematic uncertainties on the SM background for the threshold mass $\mth = 5$\,\tev.  The uncertainties are added in quadrature to obtain the total uncertainty.}
  \label{tab:systematics}
  \begin{center}
    \begin{tabular}{lrrcrr}
      \hline\hline\\[-0.9em]
      Source         & \multicolumn{2}{c}{Electron+jet} &~~~~~& \multicolumn{2}{c}{Muon+jet} \\
      & \multicolumn{2}{c}{\%} & &  \multicolumn{2}{c}{\%} \\
      \hline\\[-0.9em]
      Lepton reconstruction,     & ${+2}$&${-1}$    && ${+30}$&${-7}$\\
      scale and resolution      &&&&&\\
      Jet reconstruction,        & ${+31}$&${-15}$   && ${+5}$&${-5}$ \\
      scale and resolution      &&&&&\\
      Multijet modeling         & ${+27}$&${-27}$   && \multicolumn{2}{c}{-}               \\
      PDF                       & ${+52}$&${-33}$   && ${+100}$&${-69}$  \\
      Fit                       & ${+77}$&${-77}$   && ${+130}$&${-71}$ \\
      \hline\\[-0.9em]
      Total                     & ${+100}$&${-89}$   && ${+170}$&${-100}$ \\[0.1em]
      \hline\hline
    \end{tabular}
  \end{center}
\end{table}

The systematic uncertainties on the background are evaluated as a function of \minv, and are dominated by uncertainties on 
the fits used to extrapolate the background to the highest \minv,  uncertainties on PDFs,  and the choice of MC generator.
Systematic uncertainties due to the choice of fitting functions are evaluated by fitting the \minv\ spectrum with parameters $p_3$ and $p_4$ free and taking the difference between these fits and the fits with $p_3$ and $p_4$ fixed to zero. Additionally, \sherpa\ samples are used instead of \alpgen\ and the fits are repeated.
The uncertainty in the PDFs is estimated using a set of $44$ PDF eigenvectors for CTEQ6.6~\cite{cteq66}.    
For each of the $44$ sets, the background fits are repeated and the extrapolated backgrounds are estimated.  To estimate the uncertainty in the multijet background in the electron channel, an alternative selection of background-enriched data events, based on photons, is used.
The systematic uncertainties from the simulation of the detector response are associated with the jet and electron energy scales 
and resolutions, the muon momentum scale and resolution, and the trigger requirement.  The combined uncertainty in the background prediction ranges from $16$\% ($1$\,\tev) to  $100 $\% ($6$\,\tev) for the electron channel and from $50$\% ($1$\,\tev) to $170$\% ($6$\,\tev) for the muon channel.  Background systematic uncertainties for $\mth=5$\,\tev\ are given in Table~\ref{tab:systematics}.  
\begin{table}[t]
  \caption{
    Numbers of expected background (Exp.) and observed (Obs.) events, along with the cumulative signal efficiencies (Eff.), with uncertainties including both the statistical and systematic components for various values of \mth. Numbers of events are integrated above \minv\ requirement for the given \mth. } 
  \label{tab:results}
  \begin{center}
    \begin{tabular}{lrr@{}r@{$^{+}_{-}$}lcrr@{}r@{$^{+}_{-}$}lc}
      \hline\hline\\[-0.9em]
      $\mth$          & \multicolumn{5}{c}{Electron+jet}               & \multicolumn{5}{c}{Muon+jet} \\
      &  Obs. & \multicolumn{3}{c}{Exp.} & Eff.        & Obs. &  \multicolumn{3}{c}{Exp.} &Eff.  \\
      \hline\\[-0.9em]
      \tev &       &  \multicolumn{3}{c}{ }  & \%          &      &  \multicolumn{3}{c}{ }      &\%  \\
      \hline\\[-0.9em]
      1.0                     & $1200$& $1210$&&$^{230}_{220}$   & $57\pm4$  &     $620$ &  $550$&\multicolumn{2}{l}{$\pm280$}            & $38\pm4$ \\[0.2em]
      1.5                     & $100$ & $110$&\multicolumn{2}{l}{$\pm 40$}   & $57\pm4$  &     $49$  &  $65$&&$^{45}_{40}$       & $36\pm4$\\[0.2em]
      2.0                     & $12$  & $19$&&$^{13}_{12}$       & $56\pm4$  &     $8$   &  $14$&&$^{16}_{14}$       & $36\pm4$\\[0.2em] 
      2.5                     & $0$   & $5.3$&&$^{4.5}_{3.9}$    & $55\pm4$  &     $3$   &  $5$&&$^{6}_{5}$          & $34\pm4$ \\[0.2em] 
      3.0                     & $0$   & $1.8$&&$^{1.8}_{1.6}$    & $54\pm4$  &     $1$   &  $2.1$&&$^{2.9}_{2.1}$    & $34\pm4$ \\[0.2em] 
      3.5                     & $0$   & $0.76$&&$^{0.79}_{0.67}$ & $54\pm4$  &     $0$   &  $1.0$&&$^{1.6}_{1.0}$  & $33\pm4$ \\[0.2em] 
      4.0                     & $0$   & $0.35$&&$^{0.38}_{0.34}$ & $53\pm4$  &     $0$   &  $0.57$&&$^{0.94}_{0.57}$ & $33\pm5$ \\[0.2em] 
      5.0                     & $0$   & $0.09$&&$^{0.10}_{0.09}$ & $52\pm4$  &     $0$   &  $0.24$&&$^{0.39}_{0.24}$ & $32\pm5$\\[0.2em] 
      6.0                     & $0$   & $0.03$&&$^{0.04}_{0.03}$ & $52\pm4$  &     $0$   &  $0.13$&&$^{0.22}_{0.13}$ & $32\pm6$\\[0.2em] 
      \hline\hline
    \end{tabular}
  \end{center}
\end{table}

Uncertainties on the signal efficiency in each of the mass bins are associated with the requirements on $\Delta \eta$, $\Delta \phi$, $\langle \eta \rangle $, $\minv$, and isolation.  In addition, uncertainties on the detector simulation, mentioned above for the background, as well as the uncertainty in luminosity are taken into account.  The combined uncertainty in the signal efficiency from these sources ranges from 
$3.5$\% at $1$\,\tev\ to $3.9$\% at $6$\,\tev\ for the electron channel and from $3.6$\% at $1$\,\tev\ to $5.6$\% at $6$\,\tev\ for the muon channel.  The cumulative efficiency, shown in Table~\ref{tab:results}, is taken from the signal MC simulation for charge $+4/3$ QBHs.  The differences in the efficiency between the charge $+4/3$ state and the other charged states are much smaller than the uncertainties mentioned above and are neglected. 
The effect of the $0.65$\% uncertainty in the LHC beam energy~\cite{lhc:beamenergy} is to
change the QBH production cross section.  Since the QBH cross section is nearly constant
in $\mth/\sqrt{s}$ this is effectively an uncertainty in \mth\ and has a negligible effect on the limits.

The observed numbers of events and the expected backgrounds, shown in Table~\ref{tab:results}, are in  agreement within the total 
uncertainty.  There is no evidence for any excess.  Upper limits on  $\BFCS$ for the production of QBHs above \mth\ are determined in the interval $1-6$\,\tev\ assuming lepton universality and  using the CLs 
method~\cite{TJunk,Read:2002hq}, which is designed to give conservative limits in cases where the observed background fluctuates below the expected values.  The statistical combination of the channels employs a likelihood function constructed as the product of 
Poisson probability terms describing the total number of events observed in each channel. Systematic uncertainties are incorporated as nuisance parameters into the likelihood through their effect on the mean of the Poisson functions and through convolution with their assumed Gaussian distributions.  Correlations between channels are taken into account. 

Figure~\ref{fig:limits} shows the $95$\% confidence level (C.L.) combined lepton+jet upper limit on the cross section times branching 
fraction for the production of QBHs as a function of \mth.  Above $3.5$\,\tev,  the limit is $0.18$\,fb.  For the 
$n=6$ QBH model assumed in this Letter, the $95$\% C.L. lower limit on \mth\ is $5.3$\,\tev.  For $n=2$, and all other model assumptions the same, the $95$\% C.L. lower limit on \mth\ is $4.7$\,\tev.  
Treating the channels separately, the $95$\% C.L. upper limit on the electron (muon)+jet  $\BFCS$ above $3.5$\,\tev\ is $0.27$ ($0.49$)\,fb, and the $n=6$ lower limit on \mth\ is $5.2$ ($5.1$)\,\tev.  
\begin{figure}[t]
  \includegraphics[width=\columnwidth,trim=11 11 10 10,clip=true]{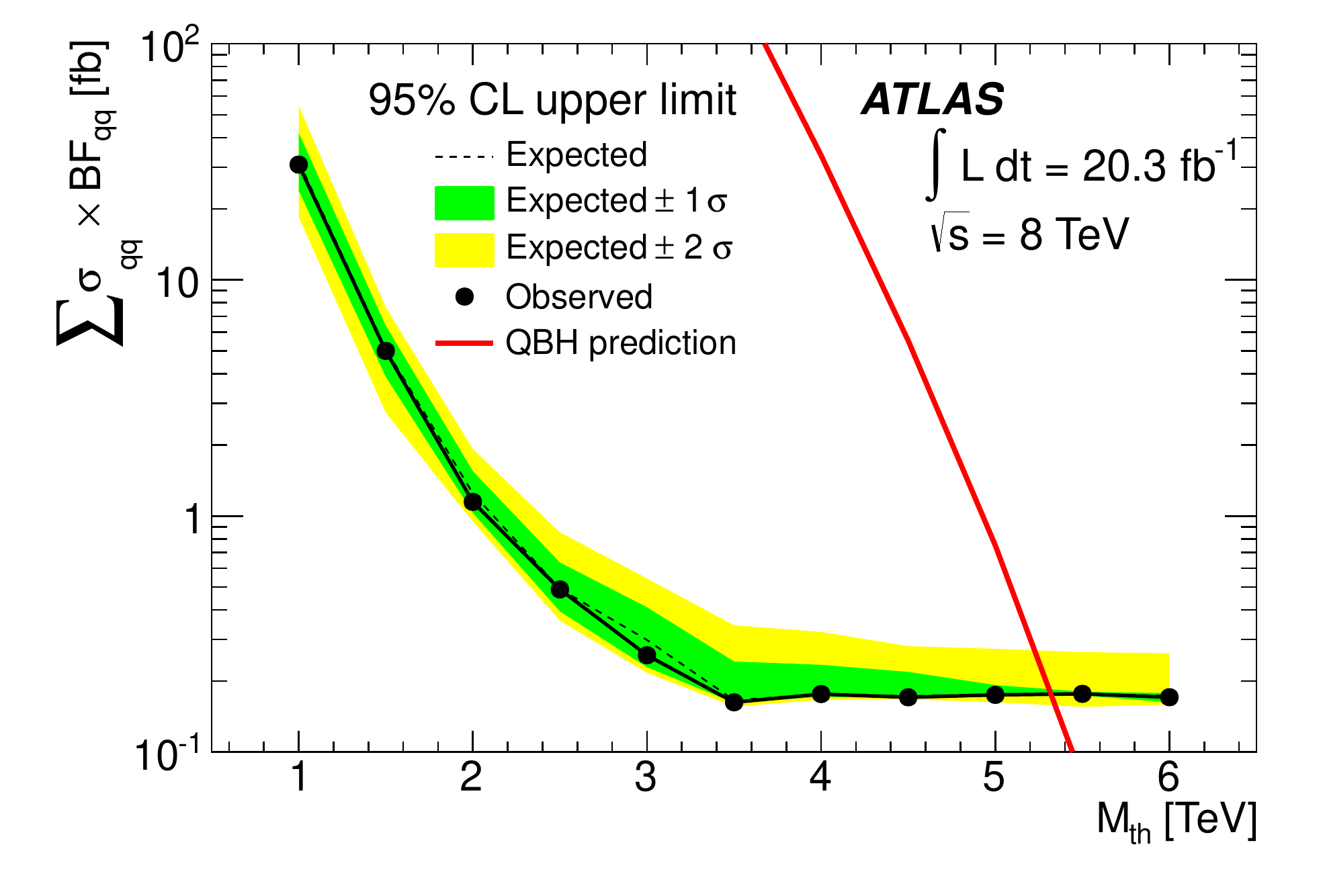}
  \caption{\label{fig:limits} The combined $95$\% C.L. upper limits on $\BFCS$ for QBHs decaying to a lepton and jet, as a function
    of \mth, assuming \md = \mth\ and $n = 6$ ADD extra dimensions. The limits take into account statistical and systematic
    uncertainties. Points along the solid black line indicate the mass of the signal where the limit is computed.  Also shown are 
    the $\pm 1 \sigma$ and  $\pm 2 \sigma$ bands indicating the underlying distribution of possible limit outcomes under the 
    background-only hypothesis.  The predicted cross section for QBHs is shown as the solid curve.}  
\end{figure}

In conclusion, a first search for two body lepton+jet final states with large invariant mass has been performed using $20.3$\,fb$^{-1}$ of $pp$ 
collisions recorded at $\sqrt{s}=8$\,\tev\ with the ATLAS detector at the LHC.  In the invariant-mass region above  $1$\,\tev\ the 
observed events are consistent with data-driven extrapolated backgrounds from the low-invariant-mass control region.  Above $3.5$\,\tev\ the expected background drops below one event and the $95$\% C.L. upper limit on the electron (muon)+jet $\BFCS$ is $0.27$ ($0.49$)\,fb.  Assuming lepton universality, the 95\% C.L. upper limit on the sum of the product of QBH lepton+jet production cross sections and branching fractions is 0.18 fb.

\begin{acknowledgments}
We thank CERN for the very successful operation of the LHC, as well as the
support staff from our institutions without whom ATLAS could not be
operated efficiently.

We acknowledge the support of ANPCyT, Argentina; YerPhI, Armenia; ARC,
Australia; BMWF and FWF, Austria; ANAS, Azerbaijan; SSTC, Belarus; CNPq and FAPESP,
Brazil; NSERC, NRC and CFI, Canada; CERN; CONICYT, Chile; CAS, MOST and NSFC,
China; COLCIENCIAS, Colombia; MSMT CR, MPO CR and VSC CR, Czech Republic;
DNRF, DNSRC and Lundbeck Foundation, Denmark; EPLANET, ERC and NSRF, European Union;
IN2P3-CNRS, CEA-DSM/IRFU, France; GNSF, Georgia; BMBF, DFG, HGF, MPG and AvH
Foundation, Germany; GSRT and NSRF, Greece; ISF, MINERVA, GIF, DIP and Benoziyo Center,
Israel; INFN, Italy; MEXT and JSPS, Japan; CNRST, Morocco; FOM and NWO,
Netherlands; BRF and RCN, Norway; MNiSW, Poland; GRICES and FCT, Portugal; MNE/IFA, Romania; MES of Russia and ROSATOM, Russian Federation; JINR; MSTD,
Serbia; MSSR, Slovakia; ARRS and MIZ\v{S}, Slovenia; DST/NRF, South Africa;
MINECO, Spain; SRC and Wallenberg Foundation, Sweden; SER, SNSF and Cantons of
Bern and Geneva, Switzerland; NSC, Taiwan; TAEK, Turkey; STFC, the Royal
Society and Leverhulme Trust, United Kingdom; DOE and NSF, United States of
America.

The crucial computing support from all WLCG partners is acknowledged
gratefully, in particular from CERN and the ATLAS Tier-1 facilities at
TRIUMF (Canada), NDGF (Denmark, Norway, Sweden), CC-IN2P3 (France),
KIT/GridKA (Germany), INFN-CNAF (Italy), NL-T1 (Netherlands), PIC (Spain),
ASGC (Taiwan), RAL (UK) and BNL (USA) and in the Tier-2 facilities
worldwide.

\end{acknowledgments}

\bibliography{qbh2012}

\clearpage
\widetext
\begin{flushleft}
{\Large The ATLAS Collaboration}

\bigskip

G.~Aad$^{\rm 48}$,
T.~Abajyan$^{\rm 21}$,
B.~Abbott$^{\rm 112}$,
J.~Abdallah$^{\rm 152}$,
S.~Abdel~Khalek$^{\rm 116}$,
O.~Abdinov$^{\rm 11}$,
R.~Aben$^{\rm 106}$,
B.~Abi$^{\rm 113}$,
M.~Abolins$^{\rm 89}$,
O.S.~AbouZeid$^{\rm 159}$
% ...
\bigskip
\\
% ...
$^{11}$ Institute of Physics, Azerbaijan Academy of Sciences, Baku, Azerbaijan\\
% ...
$^{21}$ Physikalisches Institut, University of Bonn, Bonn, Germany\\
% ...
$^{48}$ Fakult{\"a}t f{\"u}r Mathematik und Physik, Albert-Ludwigs-Universit{\"a}t, Freiburg, Germany\\
% ...
$^{89}$ Department of Physics and Astronomy, Michigan State University, East Lansing MI, United States of America\\
% ...
$^{106}$ Nikhef National Institute for Subatomic Physics and University of Amsterdam, Amsterdam, Netherlands\\
% ...
$^{112}$ Homer L. Dodge Department of Physics and Astronomy, University of Oklahoma, Norman OK, United States of America\\
$^{113}$ Department of Physics, Oklahoma State University, Stillwater OK, United States of America\\
% ...
$^{116}$ LAL, Universit{\'e} Paris-Sud and CNRS/IN2P3, Orsay, France\\
% ...
$^{152}$ Institute of Physics, Academia Sinica, Taipei, Taiwan\\
% ...
$^{159}$ Department of Physics, University of Toronto, Toronto ON, Canada\\
% ...
\end{flushleft}

\end{document}